\documentclass[pre, twocolumn]{revtex4}
\usepackage{amsmath}
\usepackage{lipsum}
\usepackage{braket}
\usepackage{bm}
\usepackage[dvipdfmx]{color}
\usepackage[dvipdfmx]{graphicx}

\newcommand{\ignore}[1]{}

\begin{document}
\title{Study on Effects of Dipole-dipole Interactions in Nd$_{2}$Fe$_{14}$B thin film Based on Stochastic Cut-off Method with New Efficient Algorithm }
\author{Taichi Hinokihara$^{1,2,*}$}
\email{hinokihara@spin.phys.s.u-tokyo.ac.jp}
\author{Masamichi Nishino$^{2,3}$}
\author{Yuta Toga$^{2}$}
\author{Seiji Miyashita$^{1,2}$}
\affiliation{%
$^1$Department of Physics, Graduate School of Science,
The University of Tokyo, 7-3-1 Hongo, Bunkyo-Ku, Tokyo 113-8656, Japan\\
$^2$Elements Strategy Initiative Center for Magnetic Materials(ESICMM), National Institute for Materials Science, Tsukuba, Ibaraki, Japan\\
$^3$International Center for Materials Nanoarchitectonics, National Institute for Materials Science, Tsukuba, Ibaraki, Japan\\
}
\begin{abstract}
	We developed a new algorithm for the stochastic cutoff method, which is a Monte Carlo method for long-range interacting systems.
 	The present new algorithm is efficient for complicated lattice structures, e.g., amorphous and also materials with complicated unit cell structure.
	In addition, it has an advantage in a high temperature region.
	With the present method, we investigated an atomistic model of the permanent magnet $\mathrm{Nd_{2}Fe_{14} B}$ with dipole-dipole interactions
	to study the effect of the anisotropy of the Fe ions on the magnetization structure in the thin film system.
	It has revealed that the anisotropy of the Fe ions plays an important role to maintain the uniaxial magnetization structure at high temperatures.
\end{abstract}
\maketitle
\section{Introduction}
Study on manufacturing magnets with high coercive force has attracted much attention in various fields of science and technology~\cite{Herbst1991}.
In the development of magnetic materials, the discovery of $\mathrm{Nd_{2}Fe_{14} B}$ has given a huge impact~\cite{Sagawa1984,Croat1984}.
It exhibits the strongest magnetic performance and has been used in a variety of applications.
Further improvements of the performance of this magnet at high temperatures are particularly desirable for industrial applications~\cite{Vial2002,Li2009}.
For this purpose, it is important to clarify the thermal effects and the dominant mechanism of the magnetization reversal process of this magnet.
Thus, theoretical studies based on an atomistic spin model, in which the magnetic moment at each atom is taken into account, is required.

Recently, Toga et al. have proposed the atomistic spin model for $\mathrm{Nd_{2}Fe_{14}B}$~\cite{Toga2016, Nishino2017}, in which the exchange couplings, magnetic anisotropies, and spin moments are derived from the first-principles calculation~\cite{Miura2014} and the experimental results~\cite{Yamada1988}.
In addition, they have also evaluated this model using the Monte Carlo simulation and have confirmed that the magnetic properties, such as temperature dependence of magnetization, anisotropy energy and so on, quantitatively agree with experimental results~\cite{Toga2016}.
The temperature dependence of the domain walls has been also studied in this model~\cite{Nishino2017,Hirosawa2017}.
Here, we note that the effect of the dipole-dipole interaction (DDI) are not taken into account in their calculations.

In general, the large magnetic anisotropy of $J=9/2$ on the Nd ions is considered to be a dominant factor of the high coercivity of this magnet.
However, it has been pointed out that their effects almost disappear owing to thermal fluctuation in a high temperature region,
while the anisotropy on the Fe ions remains in this region~\cite{Skomski1998a,Sasaki2015,Toga2016}.
To confirm this claim, it is worth investigating how the magnetic structure of the thin film changes with DDI between atomistic models with and without the anisotropy on the Fe ions.
It is expected that all the spins are oriented to the in-plane axis by introducing the DDI when the magnetic anisotropy is negligibly small compared with the DDI.
Therefore, introducing the DDI enables us to evaluate the effect of anisotropy on the Fe ions.

Although the DDI plays an significant role for physical properties in magnets~\cite{Frei1957,Booth1995,Iwano2015},
numerical simulations on long-range interacting systems are one of the challenging subjects for computational physics.
As an example, the naive Monte Carlo (MC) simulation on those systems costs $\mathcal{O}(N^2)$ computational time per one MC step (MCS), where $N$ is the number of spins.
In order to reduce the computational time, several methods have been proposed~\cite{Appel1985,Barnes1986,Carrier1988,Makino1990,Darden1993,Essmann1995,LUIJTEN1995,Hinzke2000}.
In particular, two efficient methods to calculate the equilibrium state without approximations have been proposed: the stochastically cutoff (SCO) method proposed by Sasaki and Matsubara~\cite{Sasaki2008,Sasaki2010} and the order-$N$ cluster MC method by Fukui and Todo~\cite{Fukui2009}.
These methods can calculate long-range interacting systems in reasonable computational time.
As an example, one MCS in the three-dimensional system with the DDI can be computed in $\mathcal{O}\left(\beta N\log N\right)$ by using the SCO method, where $\beta$ is the inverse temperature.
Since the SCO method can treat the classical spin system straightforwardly, we employ this method to simulate the atomistic spin model.
Moreover, it should be noted that the SCO method can parallelize the MC simulation in spite of the long-range interacting system~\cite{Endo2015a}. 

The SCO method is based on the idea of the stochastic potential switching (SPS) algorithm~\cite{Mak2005a}.
In this method, all the DDI are stochastically switched either to zero or to pseudo interactions before updating spins.
Although computing all these switching one-by-one costs $\mathcal{O}\left(\beta N^2\right)$ computational time,
Sasaki and Matsubara have overcome this difficulty by implementing a kind of inverse function method with classifying all the bonds to groups according to the type of bonds.

Their algorithm is efficient when the number of bonds stored in each group is large.
For example, a system with the translational invariance, the number of bonds included in a group is roughly proportional to the number of spins in the system.
In other words, different algorithms are required for systems with complicated structures, such as amorphous systems, junction systems, and systems including impurity spins.
Incidentally, the atomistic spin model for $\mathrm{Nd_{2}Fe_{14} B}$ contains 68 spins for each unit cell~\cite{Toga2016, Nishino2017}, which is almost the amorphous system.

In this paper, we propose a new computational method for the SPS algorithm.
The present method is inspired by the numerical method implemented in the order-$N$ cluster MC method proposed by Fukui and Todo~\cite{Fukui2009}.
Their mehtod has the advantage that it can be applied regardless of symmetries in a system. 
Thus, we can evaluate the effect of the DDI on systems with complicated structures, such as the atomistic spin model of $\mathrm{Nd_{2}Fe_{14} B}$, by using the present method.

This paper is organized as follows.
In \S 2, we introduce the atomistic spin model of $\mathrm{Nd_{2}Fe_{14} B}$ with the DDI.
In \S 3, we introduce the SCO method using the new algorithm.
The efficiency of the present method based on the atomistic spin model is also discussed in comparison with the original SCO method proposed by Sasaki and Matsubara.
In \S 4, we show the numerical results for the $\mathrm{Nd_{2}Fe_{14} B}$ thin film.
In \S 5. the summary and discussion are given.

\section{Atomistic model Hamiltonian for $\mathrm{Nd_2Fe_{14}B}$ }
\begin{figure}[t]
\centering
\includegraphics[width=\hsize,keepaspectratio]{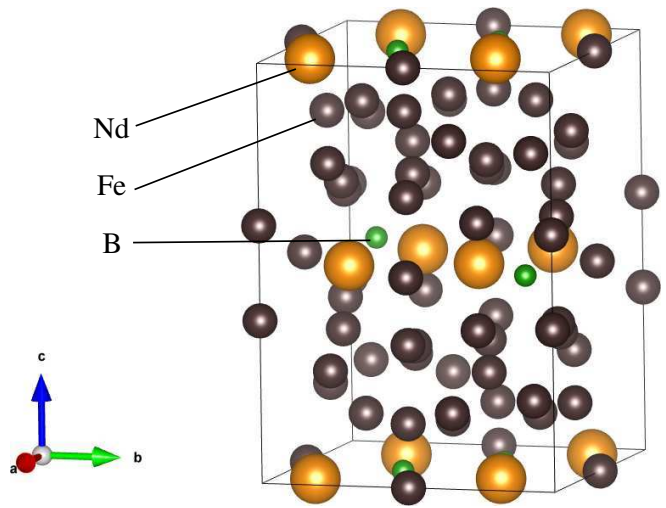}
\caption{ 
	Crystal structure of $\mathrm{Nd}_{2}\mathrm{Fe}_{14}\mathrm{B}$ magnet.
	This figure was plotted using VESTA~\cite{Momma2011}.
}
\label{Fig:NdFeB}
\end{figure}

In this section, we briefly introduce the atomistic Hamiltonian of $\mathrm{Nd_2Fe_{14}B}$ with DDIs.
Figure~\ref{Fig:NdFeB} shows the unit cell of $\mathrm{Nd_2Fe_{14}B}$, where the lattice constants for $a$-, $b$-, and $c$-axis are $d_a = d_b=8.8$\AA\,and $d_c = 12.19$\AA, respectively.
There are 68 atoms occupying nine crystallographically inequivalent sites in the unit cell.
The atomistic Hamiltonian was proposed by Toga, et al.~\cite{Toga2016}, which is given in the following form:
\begin{align}
	\mathcal{H}_\mathrm{atom} =& - \sum_{l} 2 J^{\mathrm{ex}}_{l} \bm{s}_{i_l}\cdot\bm{s}_{j_l}\nonumber\\
	&- \sum_{i\in \mathrm{Fe}}K^{\mathrm{Fe}}_i\left(s_i^{z}\right)^2
	- \sum_{i\in \mathrm{Nd}} \sum_{l=2,4,6} K^{\mathrm{Nd}(l)}_{i} s_i^{zl} \nonumber\\
	&-\sum_i S_i\bm{h}\cdot \bm{s}_i.
	\label{Hamiltonian_atom}
\end{align}
Here, $l$ is the bond index, $i_l$ and $j_l$ denote site indices connected with the $l$-th bond,
$\bm{s}_i$ and $S_i$ are the normalized spin vector ($|\bm{s}_i|=1$) and the magnitude of the magnetic moment as a unit of Bohr magneton $\mu_B = 9.27410\times 10^{-24}\mathrm{J}/ \mathrm{T}$ on $i$th site, respectively,
and $\bm{h}$ is the external magnetic field expressed in a proper unit.

It should be noted that Eq. \eqref{Hamiltonian_atom} treats the 5d and 6s electrons and the 4f electrons ($J=9/2$) on the Nd ions separately. 
The anisotropy is given for the spin representing $J=9/2$, while the interaction with surrounding spins are attributed to the spin representing the 5d and 6s electrons.
These two spins are assumed to be completely antiparallel in the present model.
$K^{\mathrm{Nd}(l)}_{i}$ ($l=2$, 4, and 6) represents the anisotropy of $J=9/2$ on the Nd ions and are determined from the experimental results~\cite{Yamada1988}.
The amplitude of these values are $K^{\mathrm{Nd}(2)} = -3.57\times10^2\mathrm{K}$, $K^{\mathrm{Nd}(4)} = 1.88\times10^3\mathrm{K}$, and $K^{\mathrm{Nd}(6)} = -1.66\times10^3\mathrm{K}$, respectively~\cite{Toga2016}.
Owing to the higher-order anisotropy, the easy axis of this spin inclines at $36.7^\circ$ from the $c$-axis.
Here, the anisotropy for the 5d and 6s electrons on the Nd ions is assumed to be 0.
On the other hand, the anisotropy of the Fe ions $K^{\mathrm{Fe}}_i$ are assumed to be uniaxial, derived from the first-principles density-functional calculation for $\mathrm{Y_2Fe_{14}B}$~\cite{Miura2014}.

All the exchange couplings $J^{\mathrm{ex}}_{l}$ were evaluated from the first-principles calculation using Lichtenstein's formula on the Korringa-Kohn-Rostoker (KKR) Green's-function method~\cite{Liechtenstein1987,Toga2016}.
We set the cutoff range for exchange couplings 3.52\AA\ in the same way as the previous work~\cite{Nishino2017},
in which all dominant short-range interactions including Nd-Fe interactions are taken into account.

The DDI is introduced to this system as,
\begin{align}
	\mathcal{H} =& \mathcal{H}_{\mathrm{atom}}
	+ \sum_{l,\alpha,\alpha^\prime} D^{\alpha\alpha^\prime}_{l} s^{\alpha}_{i_l} s^{\alpha^\prime}_{j_l},
	\label{Hamiltonian}
\end{align}
where $\alpha$ and $\alpha^\prime$ denote the $x$, $y$, or $z$ coordinate;
$D_l^{\alpha\alpha^\prime}$ is the amplitude of the DDI of the $\alpha\alpha^\prime$ component.
The number of bonds for the DDI $N_b$ is of the order of $\mathcal{O}\left(N^2\right)$.
Once a system structure and all the $S_i$ are given, all the $D^{\alpha\alpha^\prime}_l$ are given by
\begin{align}
	D_l^{\alpha\alpha^\prime} = 
	\begin{cases}
		\frac{\mu_0}{4\pi}\left(\frac{1}{r_{l}^3}- 3\frac{r_{l\alpha}r_{l\alpha^\prime}}{r_{l}^5}\right)S_{l_i}S_{j_l}& \alpha =\alpha^\prime\\
		-3\frac{\mu_0}{4\pi}\frac{r_{l\alpha}r_{l\alpha^\prime}}{r_{l}^5}S_{l_i}S_{j_l}& \alpha \neq \alpha^\prime
	\end{cases}.
\end{align}
Here, $\bm{r}_l$ represents the distance vector of the $l$-th bond
and $\mu_0$ is the vacuum magnetic permeability.

\section{SCO Method}
In order to calculate the system written in Eq.~\eqref{Hamiltonian} within a reasonable computational time, we apply the SCO method to the DDI.
Other terms, i.e., exchange couplings and anisotropies, do not perform the switching of bonds and are always treated in original forms at the spin update process.

Let us briefly introduce the SCO method proposed by Sasaki and Matsubara~\cite{Sasaki2008}.
Hereafter, we represent the $l$-th term of the DDI as $V_l\left(x\right)$:
\begin{align}
	V_l\left(x\right) \equiv \sum_{\alpha\alpha^\prime} D^{\alpha\alpha^\prime}_l s^{\alpha}_{i_l}s^{\alpha^\prime}_{j_l},
	\label{dd_int}
\end{align}
where $x$ represents the configuration of all the spins $\{\bm{s}_{1},\bm{s}_{2},\ldots\}$ in the system.
It is obvious that the amplitude of $V_l\left(x\right)$ only depends on the $i_l$-th and $j_l$-th spins, although we use $x$ for simplicity.

The key ingredient of the SCO method is the bond switching procedure: 
the $l$-th bond $V_{l}\left(x\right)$ is stochastically replaced by either to 0 with the probability $p_l\left(x\right)$ or to the pseudo-interaction $\overline{V}_l\left(x\right)$ with the probability $1-p_l\left(x\right)$.
These values are determined so as to hold the original detailed balance condition:
\begin{align}
	e^{-\beta\mathcal{H}\left(x\right)} W\left(x \rightarrow x^\prime \right) = e^{-\beta\mathcal{H}\left(x^\prime\right)} W\left(x^\prime \rightarrow x \right),
	\label{original_detail}
\end{align}
where $W\left(x \rightarrow x^\prime \right)$ is the transition probability from $x$ to $x^\prime$,
and $\mathcal{H}\left(x\right)$ is the total energy of the configuration $x$.
Hereafter, we call the case in which $\overline{V}_l\left(x\right)$ is selected \lq\lq accepted", while we call the other case \lq\lq rejected".
$\overline{V}_l\left(x\right)$ and $p_l\left(x\right)$ for all the bonds can be derived as follows.

First, we discuss the switching of the $i$-th bond, where all the other bonds are treated in the original form $V_l\left(x\right)$.
The bond switching with probability $p_i\left(x\right)$ transforms the original transition probability $W\left(x \rightarrow x^\prime \right)$ into the following form:
\begin{align}
	W\left(x \rightarrow x^\prime \right) = p_i\left(x\right) \tilde{W}\left(x \rightarrow x^\prime \right) + \left(1-p_i\left(x\right)\right) \overline{W}\left(x \rightarrow x^\prime \right),
	\label{trans_SCO}
\end{align}
where $\tilde{W}\left(x \rightarrow x^\prime \right)$ ($\overline{W}\left(x \rightarrow x^\prime \right)$) is the transition probability of the system in which the $i$-th bond is rejected (accepted).
In order to hold Eq.~(\ref{original_detail}), $\overline{V}_{i}\left(x\right)$ and $p_{i}\left(x\right)$ are given by,
\begin{align}
	\label{eq:pseudo_p}
	\overline{V}_{i}\left(x\right) = V_{i}\left(x\right) - \beta^{-1}\log\left[1 -  e^{\beta\left(V_i\left(x\right) - V_i^\ast\right)}\right],\\
	\label{eq:probability}
	p_i\left(x\right) = \exp\left[\beta\left(V_i\left(x\right) - V_i^\ast\right)\right].
\end{align}
Here, $V_i^{\ast}$ is a constant greater than or equal to the maximum value of $V_i\left(x\right)$.
Although the dynamics of spins depends on $V^{\ast}_i$, the equilibrium state does not change by the value.
In the following discussion, we assume $V_i^{\ast} = \mathrm{max}[V_i\left(x\right)]$.

Second, let us discuss the switching of the $i^\prime$-th bond ($i\neq i^\prime$) after the switching of the $i$-th bond.
This switching can be executed by the same transformation 
of Eq.~(\ref{trans_SCO}) for both the $\tilde{W}\left(x \rightarrow x^\prime \right)$ and $\overline{W}\left(x \rightarrow x^\prime \right)$.
Then, it is easy to check that $\overline{V}_{i^\prime}\left(x\right)$ and $p_{i^\prime}\left(x\right)$ are expressed in the same form of Eqs.~(\ref{eq:pseudo_p}) and (\ref{eq:probability}), respectively.
Note that these expressions do not depend on whether the $i$-th bond was accepted/rejected. 
Therefore, by executing the same transformation of Eq.~(\ref{trans_SCO}) recursively,
we can prove that $\overline{V}_{l}\left(x\right)$ and $p_{l}\left(x\right)$ for all the bonds are given by the same form of Eqs.~(\ref{eq:pseudo_p}) and (\ref{eq:probability}), respectively.

After the bond switching procedure, one of the bond configurations $\Lambda$ is obtained with the probability $P_{\Lambda}$.
Here, $\Lambda$ is written as,
\begin{align}
	\Lambda \equiv \{\lambda_1, \lambda_2, \ldots \lambda_{N_b}\},\\
	\lambda_l =  
	\begin{cases}
		1 & \mathrm{: accepted} \\
		0 & \mathrm{: rejected} \\
	\end{cases},
\end{align}
and the probability $P_{\Lambda}$ is given by,
\begin{align}
	P_{\Lambda} = \prod_{l} p_l(x)^{1-\lambda_l} (1-p_l(x))^{\lambda_l}.
\end{align}

The SCO method performs the spin update procedure by using the Hamiltonian $\mathcal{H}_{\mathrm{sw}}\left(\Lambda\right)$ for the bond configuration $\Lambda$.
Since this method is a different realization of the detailed balance condition,
the correct equilibrium state is obtained as its stationary state by performing the bond switching procedure and the spin update procedure based on $\mathcal{H}_{\mathrm{sw}}\left(\Lambda\right)$ alternately.

For the bonds of $\beta V_l\ll 1$, most of them are rejected in $\mathcal{H}_{\mathrm{sw}}\left(\Lambda\right)$ owing to the small acceptance probability.
Thus, the spin update based on $\mathcal{H}_{\mathrm{sw}}\left(\Lambda\right)$ can drastically reduce the computational time compared with the naive MC simulation.
However, in principle, the bond switching procedure costs $\mathcal{O}(N^2)$ computational time when we perform the switching one-by-one.
Thus, it is required to adopt an efficient algorithm for this procedure.

As a first step, to avoid computing all the $p_l\left(x\right)$ for each bond switching procedure, we use the probability $p_l$, which is the minimum probability of $p_l\left(x\right)$ within all the spin configurations $x$.
Namely, we first reduce the candidates of accepted bonds by using $p_l$.
Here, $p_l$ is written as
\begin{align}
	p_l &= e^{-\beta\zeta_l},
\end{align}
where
\begin{align}
	\zeta_l &=  - \mathrm{min}\left[V_l\left(x\right)\right] + V_l^\ast.
\end{align}
After this preselection procedure, bonds are accepted with the probability $\left(1-p_l\left(x\right)\right)/\left(1-p_l\right)$ from among the candidates.
Since $p_l$ does not depend on the spin configuration $x$, we can store the list of $p_l$ before starting the simulation. 
Thus, this process enables us to avoid computing all the $p_l\left(x\right)$, which costs $\mathcal{O}\left(N^2\right)$ computational time, for each bond switching procedure.

The SCO method proceeds as follows:
\begin{itemize}
\item[(i)] Pick up candidates from all the bonds by using $1-p_l$. 
\item[(ii)] Accept the bonds from the candidates with probability $\left(1-p_l\left(x\right)\right)/\left(1-p_l\right)$.
\item[(iii)] Perform single spin update for all the spins using $\mathcal{H}_{\mathrm{sw}}$ for $n_{\mathrm{sw}}$ times.
\item[(iv)] Return to (i).\\
\end{itemize}
In contrast to the steps (ii) and (iii), the step (i) still costs $\mathcal{O}\left(N^2\right)$ computational time when we perform the preselection of the bond one-by-one.
Sasaki and Matsubara have overcome this difficulty by implementing a kind of inverse function method for the step (i)~\cite{Sasaki2008}.
Namely, the SCO method first constructs lists of bonds that have the same $p_l$, and then uses the inverse function method for each list~\cite{Sasaki2008}.
Note that we call one sweep of the spin update one MCS. 

The efficiency of this algorithm depends on the number of the lists.
In an extreme case, where all the $p_l$ is different with each other, the computational time for the step (i) costs $\mathcal{O}\left(N^2\right)$ because the number of lists is $N^2$.
Similar situations arise at amorphous systems, junction systems, and systems including impurity spins. 
It is conceivable that the simulation of the atomistic spin model for $\mathrm{Nd_{2}Fe_{14} B}$ also needs a large number of lists since this model contains 68 spins in the unit cell.

Therefore, to overcome this difficulty, we propose a new algorithm for the step (i).
The present idea is inspired by Fukui and Todo's idea which was used in their paper for the cluster MC method~\cite{Fukui2009}.
We call the present method as modified SCO (MSCO) method.

\subsection{MSCO method} 
In the MSCO method, we rewritten the rejection probability $p_l$ and the acceptance probability $1-p_l$
in terms of the Poisson distribution $f_{k}\left(\zeta\right)=e^{-\zeta} \zeta^k / k!$ as,
\begin{align}
	p_l &= f_0\left(\beta\zeta_l\right),\\
	1-p_l &=  \sum_{k_l=1}^\infty  f_{k_l}\left(\beta\zeta_l\right).
\end{align}
Here, for each $l$-th bond, accepted or rejected can be distinguished by $k_l$, i.e., rejected in $k_l=0$, otherwise accepted.
Although the above expression seems to be redundant, an efficient algorithm can be constructed by using properties of the Poisson distribution.

We express the bond configuration as $K=\left\{k_1,k_2\ldots\right\}$ instead of $\Lambda$, where $k_l$ is a positive integer or zero generated by the Poisson distribution with the weight of $\zeta_l$.
The acceptance probability $P_K$ for the bond configuration $K$ is written as multiplication of $f_{k}\left(\beta\zeta_l\right)$ of all the bonds and can be deformed as follows:
\begin{align}
	P_K &= \prod_l f_{k_l}\left(\beta\zeta_l\right) \nonumber\\
	&= f_{k_{\mathrm{tot}}}\left(\beta\zeta_{\mathrm{tot}}\right)
	\frac{k_{\mathrm{tot}}!}{k_1!k_2!\cdots k_{N_b}!}\prod_{l=1}^{N_b} 
	\left( \frac{\zeta_l}{\zeta_{\mathrm{tot}}} \right)^{k_l},
	\label{P_k}
\end{align}
where $\zeta_\mathrm{tot} = \sum_{l}\zeta_l$ and $k_\mathrm{tot} = \sum_{l}k_l$.
Here, we use properties of the Poisson distribution.
The second line of Eq.~(\ref{P_k}) indicates that the step (i) can be divided into following sub-steps:
\begin{itemize}
	\item[(i-a)] Set the initial list $\{k_1 = 0,k_2=0,\dots\}$.
	\item[(i-b)] Generate a random number $k_\mathrm{tot}$ from the Poisson distribution $f_{k_\mathrm{tot}}\left(\beta\zeta_\mathrm{tot}\right)$.
	\item[(i-c)] Select candidates stochastically in accordance with the weight $\zeta_l/\zeta_{\mathrm{tot}}$.
	\item[(i-d)] Add 1 to $k_l$ in the list, where $l$-th bond is selected in the sub-step (i-c).
	\item[(i-e)] Return to the sub-step (i-c) until $\sum_l k_l = k_{\mathrm{tot}}$ .
\end{itemize}
It is obvious that the sub-steps (i-b) and (i-d) only costs $\mathcal{O}\left(1\right)$ computational time, 
and the sub-step (i-c) can be also calculated in an $\mathcal{O}\left(1\right)$ computational time by using the Walker's method of alias as discussed in the appendix.
Thus, the total computational time for the step (i) is $\mathcal{O}\left(k_{\mathrm{tot}}\right)$ by using this algorithm.

The MSCO method is efficient even in amorphous systems because the classification of bonds are not necessary in contrast to the SCO method.
Moreover, since the amplitude of $k_{\mathrm{tot}}$ depends on $\beta\zeta_{\mathrm{tot}}$, this method is also efficient in a high temperature region.
On the other hand, at low temperatures, in which $k_{\mathrm{tot}}$ is larger than $N^2$, this method is inefficient because a same bond may be selected repeatedly.

\subsection{Comparison with two algorithms based on $\mathrm{Nd_2Fe_{14}B}$}
Let us compare the above two methods, SCO method and MSCO method, based on the atomistic model of $\mathrm{Nd_2Fe_{14}B}$ with the DDI.
Hereafter, the sampling and the bond switching in the both methods were performed for every 10 MCS, i.e., we set $n_{\rm sw}=10$.
Note that one MCS is defined as one sweep of the spin update for all the spins.

Throughout this paper, the magnetization behavior of the system is evaluated by the magnetization parallel to the $c$-axis $\overline{m}_z$ and by that perpendicular to the $c$-axis $\overline{m}_{xy}$, which are defined in the following forms:
\begin{align}
\overline{m}_z \equiv \sqrt{\langle m_z^2\rangle}/N,\\
\overline{m}_{xy}\equiv  \sqrt{\langle m_x^2\rangle+\langle m_y^2\rangle}/N,
\end{align}
where,
\begin{align}
m_{\alpha}=\sum_{i: {\rm all\,\,atoms}}S_i^{\alpha},\quad \alpha=x,y, z.
\end{align}

\begin{figure}[t]
\centering
\includegraphics[width=\hsize,keepaspectratio]{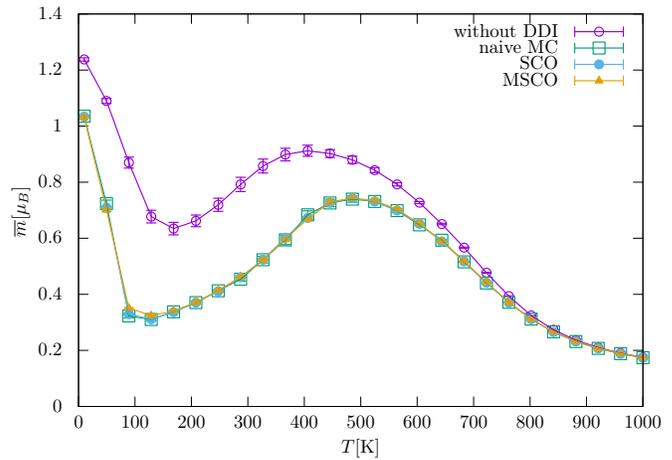}
\caption{Temperature dependence of $\overline{m}_{xy}$ for the $2\times2\times2$ unit cells system with (violet line) and without 100 times greater DDI (others).
The system with the overestimated DDI is simulated by using three methods: the naive MC (blue line), the SCO (yellow line), and the MSCO (green line).
}
\label{comparison}
\end{figure}

First, we present that the MSCO method gives the correct results obtained from the naive MC simulation.
Although systems with a small number of spins can be simulated by using the naive MC simulation,
the effect of DDI on the magnetization is difficult to see in such small-scale systems because the realistic values of the DDI are less than 1K.
Thus, we assume the amplitude of the DDI is 100 times greater than that of the original one for comparison.
Figure~\ref{comparison} shows the temperature dependence of $\overline{m}_{xy}$ with and without DDI by using three different methods: the naive MC method, SCO method, and MSCO method.
We evaluate the $2\times2\times2$ unit cells system with the open boundaries ($N=616$).
Note that the system also includes the spins located on the surface of the unit cells system.
The naive MC simulation with DDI performs 200000 MCS for equilibration and 200000 MCS for sampling,
while all the others perform 2000000 MCS for equilibration and 2000000 MCS for sampling.
We averaged the results of 10 different runs with different initial configurations.

The difference between the systems with and without the DDI clearly indicates that the magnetization curve is modified by the overestimated DDI.
Thus, we conclude that the MSCO method gives the same results of the naive MC and the SCO method as seen in Fig.~\ref{comparison}.
In the following discussion, we do not use the overestimated DDI but use the original one.

\begin{figure}[t]
\centering
\includegraphics[width=\hsize,keepaspectratio]{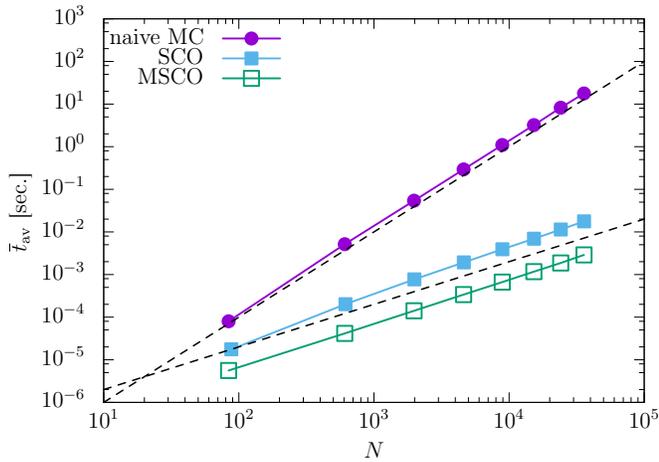}
\caption{Average computational time per one MCS at $T=400\mathrm{K}$. The calculation has done for the atomistic model of Nd-Fe-B magnet.
	     The naive calculation (violet line), the SCO method with the algorithm proposed by Sasaki and Matsubara (blue line), and the SCO method with that proposed in this paper (green line) are depicted, respectively.
	 	 Broken lines proportional to $N^2$ and $N$, respectively}
\label{benchmark}
\end{figure}

Next, we compare the computational time of these three methods.
Figure~\ref{benchmark} shows the average computational time $\overline{t}_{\mathrm{av}}$ as a function of the number of spins $N$.
While $\overline{t}_{\mathrm{av}}$ for the naive MC denotes the computational time for one MCS, that for the SCO and the MSCO methods are defined as follows:
\begin{align}
	\overline{t}_{\mathrm{av}} = t_{\mathrm{MC}}  + \frac{t_{\mathrm{sw}}}{n_{\mathrm{sw}}}
	\label{eq:tmc}
\end{align}
where $t_{\mathrm{MC}}$ and $t_{\mathrm{sw}}$ denote the computational time of one MCS and of the bond switching procedure, respectively.
We evaluate the computational time for systems from $1\times 1\times 1$ unit cell ($N=88$) to $8\times 8\times 8$ unit cells ($N=35872$) with open boundaries at 400K.
It is found that the naive MC costs $\mathcal{O}\left(N^2\right)$ computational time, while both of the SCO methods cost $\mathcal{O}\left(N\ln N\right)$ computational time.
In addition, it is also found that the MSCO method is about three times faster than the SCO method at 400K.
The following two reasons support this results.
First, since the atomistic model contains 68 spins in a unit cell, the number of bonds with different $p_l$ becomes large in this system.
Second, the DDI in this system is of the order of 1K, which is quite smaller than 400K: $k_{\mathrm{tot}} \ll N$ in the MSCO method.

\begin{figure}[t]
\centering
\includegraphics[width=\hsize,keepaspectratio]{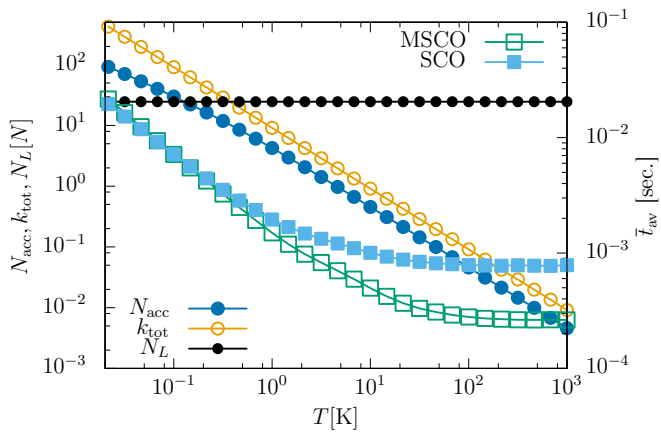}
\caption{Temperature dependence of average computational time $\overline{t}_{\mathrm{av}}$ in the case of $3\times3\times3$ unit cells system with open boundaries. 
The SCO method (blue line) and the MSCO method (green line) are depicted, respectively.
The number of acceptance bonds $N_{\mathrm{acc}}$ (dark blue line), the average of the random number generated by Poisson distribution $k_{\mathrm{tot}}$ (yellow line), and the number of lists required in the SCO method $N_L$ (black line) are also indicated, where the unit of these values are the number of spins $N$.
}
\label{benchmark_temp}
\end{figure}

To confirm these points, Fig.~\ref{benchmark_temp} depicts temperature dependence of the average computational time $\overline{t}_{\mathrm{av}}$,
the number of acceptance bonds $N_{\mathrm{acc}}$, the average of the random number generated by Poisson distribution $k_{\mathrm{tot}}$, and the number of lists required in the SCO method $N_L$.
These values are evaluated in the case of the $3\times3\times3$ unit cells system ($N=1992$) with open boundaries.
Although the number of lists $N_L$ is roughly estimated as $N/3$ in the case of simple cubic system, the present atomistic model requires approximately $25N$ lists.
This fact increases $t_{\mathrm{sw}}$ in the SCO method.
On the other hand, in the MSCO method, $k_{\mathrm{tot}}$ is smaller than the number of spins $N$ in the high temperature region, which indicates $t_{\mathrm{MC}}> t_{\mathrm{sw}}$.
Therefore, the MSCO method is faster than the SCO method in the high temperature region. 

In the low temperature region, however, the computational time for the MSCO method rapidly increases compared with the SCO method owing to the large number of $k_{\mathrm{tot}}$.
Since we focus on the permanent magnets above the room temperature, such a low temperature region is not important for the present study.
Thus, we conclude that the MSCO method is efficient in this system.
In the following section, we evaluate the atomistic spin model by using the MSCO method.


\section{Effects of DDI on the thin layer.}

In this section, we study the ordering structure in thin layers of the atomistic spin model Eq.~(\ref{Hamiltonian_atom}) with the DDI by using the MSCO method.
Here we adopt the system of $20\times 20\times 1$ unit cells with open boundary conditions.  

\begin{figure}[t]
\centering
\includegraphics[width=\hsize,keepaspectratio]{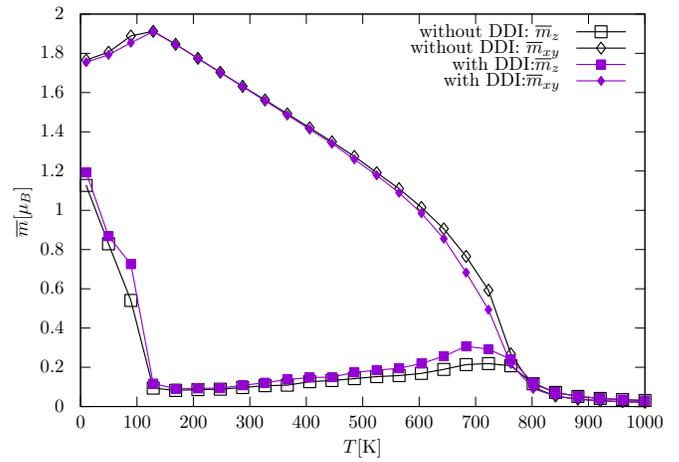}
\caption{
Temperature dependence of the magnetizations, 
$\overline{m}_z$ and $\overline{m}_{xy}$, for the atomistic model of Nd-Fe-B magnet.
	Black lines indicate the magnetization without the DDI, $\overline{m_z}$ (filled rhomboid) and $\overline{m}_{xy}$ (open rhomboid), respectively.
	On the other hand, violet lines indicate that with the DDI, $\overline{m_z}$ (filled square) and $\overline{m}_{xy}$ (open square), respectively.
	Sample size of this simulation is $20\times20\times1$.
}
\label{mt_w}
\end{figure}

In Fig.~\ref{mt_w}, we depict the temperature dependence of magnetizations $\overline{m}_z$ and $\overline{m}_{xy}$ with and without the DDI.
Both systems are simulated with 100000 MCS for equilibration and with 100000 MCS for sampling.
We averaged the results obtained from 10 different initial states.
In this sample size, we found that the DDI has little effect.

\subsection{Effect of the anisotropy of the Fe atoms}
Next we study the effect of the magnetic anisotropy of the Fe ions $K^{\mathrm{Fe}}$. 
Generally, it is considered that the strong magnetic anisotropy on the rare-earth ions is a dominant cause of the large coercive force.
However, as discussed in the previous study~\cite{Skomski1998a,Toga2016}, the anisotropy of the Fe ions seems to play an important role for the coercive force in high temperature region.
Thus, in the present paper, we study its effect on the magnetic structure under the influence of the DDI.

\begin{figure}[t]
\centering
\includegraphics[width=\hsize,keepaspectratio]{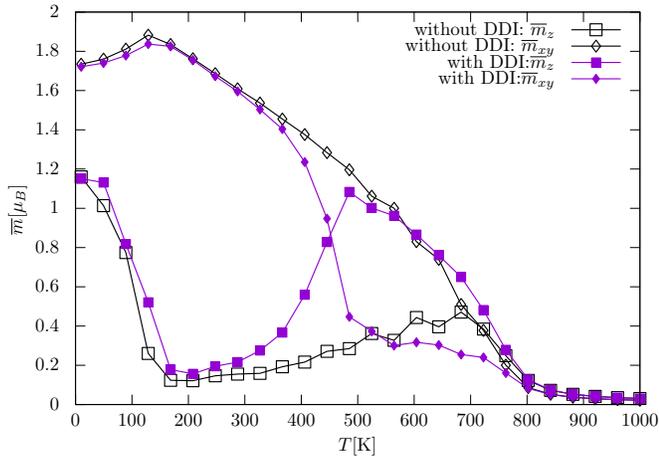}
\caption{ 
	Magnetization for the atomistic model of Nd-Fe-B magnet without anisotropy of the Fe ions as a function of temperature.
	Black lines indicate the magnetization without the DDI, $\overline{m_z}$ (filled rhomboid) and $\overline{m}_{xy} $ (open rhomboid), respectively.
	On the other hand, violet lines indicate that with the DDI, $\overline{m_z}$ (filled square) and $\overline{m}_{xy}$ (open square), respectively.
	Sample size of this simulation is $20\times20\times1$.
}
\label{mt_wo}
\end{figure}

In Fig.~\ref{mt_wo}, we depict the temperature dependence of magnetizations of the system for $K^{\rm Fe}=0$.
Here, we performed 100000 MCS for first equilibration and 100000 MCS for sampling.
We averaged the results obtained from 10 different initial states.

We found that the magnetization of the system with $K^{\rm Fe}=0$ lies in the $ab$ plane in a high temperature region (400{\rm K} $< T <$800{\rm K}), while that of real value of $K^{\rm Fe}$ stays along the $c$ axis.
Since it is well established that the DDI causes a planer structure in thin layer with small sample size,
the planer magnetic structure indicates that the effective anisotropy in the region is quite smaller than the DDI even though the anisotropy of the Nd ions is still taken into account.

This behavior is consistent with the previous study~\cite{Toga2016}, which suggested that the effective anisotropy of the Nd ions is reduced at high temperatures. 
According to Ref.~\cite{Toga2016}, the exchange coupling between the Fe-Fe ions is of the order of 230K, while that between the Fe-Nd ions is of the order of 30K.
Namely, in the high temperature region $T>400\mathrm{K}$, it is expected that the ferromagnetic state is mainly supported by the Fe ions.
In other words, the large magnetic anisotropy on the Fe ions and/or strong exchange coupling between the Nd-Fe ions would help the large coercive force at high temperatures.
This property might be evaluated in experiment by replacing Fe by Cr, Mn or Co. 

\subsection{Effect of the thickness of the layer}
\begin{figure}[t]
\centering
\includegraphics[width=\hsize,keepaspectratio]{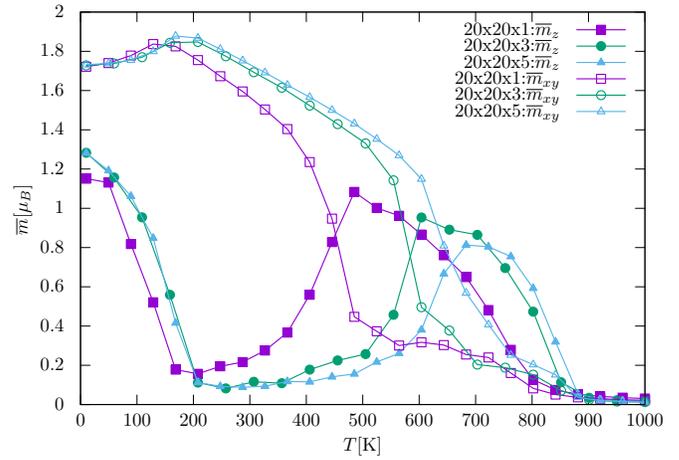}
\caption{ 
Layer dependence of magnetization for the atomistic model of Nd-Fe-B magnet without anisotropy of the Fe ions as a function of temperature.
	Sample size of this simulation is $20\times 20\times 1$ (violet lines), $20\times 20\times 3$ (green lines), and 
$20\times 20\times 5$ (light blue lines), respectively.
	Open geometries and filled geometries indicate $\overline{m}_{xy}$ and $\overline{m}_z$, respectively.
}
\label{mt_wo_l}
\end{figure}

Next, we study the effect of the thickness of the layer.
In Fig.~\ref{mt_wo_l}, we depict the dependence of magnetizations of the system of $K^{\rm Fe}=0$
on the number of layers $n_{\rm layer}$.
Here, we performed 100000 MCS for equilibration and 100000 MCS for sampling.
We averaged the results obtained from 10 different initial states.
We find that as $n_{\rm layer}$ increases, the region of the planer structure is reduced.  

This behavior can be understood as the DDI effect.
Namely, the DDI is a ferromagnetic coupling parallel to the spin direction but is an anti-ferromagnetic coupling perpendicular to the spin direction.
Thus, the anisotropy along to the $c$-axis owing to the DDI increases as increasing the number of layers while keeping the area of the system constant.

\section {Summary and Discussion}

We proposed the new algorithm of SCO method for long-range interacting systems.
The present method is efficient for systems with complicated lattice structure.
We applied this method to the atomistic spin model of $\mathrm{Nd_2Fe_{14}B}$ which has the unit cell consisting of 68 atoms.
In this case, the present algorithm is drastically reduced the computational time compared with the original algorithm proposed by Sasaki and Matsubara. 

By using the present algorithm, we discussed the effect of the anisotropy of the Fe ions on the magnetization structure at finite temperatures.
In the high temperature region, the system without the anisotropy of the Fe ions exhibits a planer structure where the DDI dominates while it has little effect at low temperatures.
Thus it was found that the anisotropy of the Fe ions plays an important role to maintain the coercive force under influence of the DDI at high temperatures. 
We also study the effect of the thickness of the layer.
It is found that the region of the planer magnetic structure is reduced as the number of layers increases.

In practical calculation, we may choose algorithms, the MSCO, the SCO, and even the naive MC concerning to their advantages, in a simulation depending on the characteristics of the bonds.
For example, we treat short-range interactions with the naive MC and use the MSCO for the DDI in the present work.
However, we may use the SCO instead of the MSCO at low temperatures.

Here, we point out that for the case of huge number of elements, 
the list used in Walker's method of alias can be separated into an arbitrarily set of lists 
thanks to the properties of the Poisson distribution.

In order to discuss the magnetic structure caused by the DDI, further large-scale system is required to be simulated.
In the present method, however, the limit of the sample size is about $100\times100\times5$ unit cells system (3400000 spins are taking into account in this system) in our simulation.
It is necessary to perform a kind of coarse-graining procedure to the atomistic Hamiltonian for simulating further large-scale systems.

\acknowledgements
We would like to thank S. Todo for useful discussions and information.
This work is supported by the Elements Strategy Initiative Center for Magnetic Materials (ESICMM) under the outsourcing project of MEXT.
The authors thank the Supercomputer Center, the Institute for Solid State Physics, The University of Tokyo, for the use of the facilities.

\appendix*
\section{Walker's method of alias in the present method}
In this appendix, we briefly explain the Walker's method of alias used in the sub-step (i-b).
The idea of this method is an extension of the simplest $\mathcal{O}\left(1\right)$ method to pick up an element form 
an ensemble of $N$ elements with the uniform probability distribution.
In this case, we can simply obtain the $i$-th element ($1 \leq i \leq N$) by generating an uniformly distributed integer random number.
Walker's method of alias extends the above idea to the case of a general probability distribution.

This method requires two tables of size $N$.
One is the table of alias numbers $\{A_i\}$  ($1 \leq A_i \leq N$) and the other is the modified probabilities ${P_i}$ ($0\leq P_i \leq 1$).
Using these tables, the Walker algorithm proceeds as follows:
\begin{itemize}
\item[(i)]Generate a uniformly distributed integer random number $i$, ($1\leq i\leq N$).
\item[(ii)]Generate a uniformly distributed real random number $r$, ($0\leq r < 1$).
\item[(iii)]If $r<P_i$, $i$-th element is chosen. Otherwise, $A_i$-th element is chosen.
\end{itemize}
Obviously, this procedure is calculated in $\mathcal{O}\left(1\right)$ computational time and is a kind of the rejection free method.

Both of the tables, $\left\{ P_i\right\}$ and $\left\{A_i\right\}$, can be systematically obtained as proposed by Fukui and Todo~\cite{Fukui2009}.
First, we prepare a table $\left\{ P^\prime_i\right\}$ where $P^\prime_i$ is the acceptance probability of the $i$-th element multiplied by $N$.
and rearrange this table so that all the elements with $P^\prime_i \geq 1$ precede with those with $P^\prime_i < 1$.
Then, we create the tables $\left\{ P_i\right\}$ and $\left\{A_i\right\}$ from the right most element to the left most element sequentially in the following procedures.
\begin{itemize}
\item[(a)] Set $A_i = i^\prime$, where $i^\prime$-th element is the right most element with $P^\prime_{i^\prime} \geq 1$.
\item[(b)] Set $P_i = P^\prime_i$.
\item[(c)] Redefine $P^\prime_{i^\prime}$ as $P^\prime_{i^\prime} - (1 - P_i) $.
\item[(d)] Proceed $i$ to the next element and return to step (a).
\end{itemize}

\bibliography{main.bbl}

\begin{thebibliography}{30}
\expandafter\ifx\csname natexlab\endcsname\relax\def\natexlab#1{#1}\fi
\expandafter\ifx\csname bibnamefont\endcsname\relax
  \def\bibnamefont#1{#1}\fi
\expandafter\ifx\csname bibfnamefont\endcsname\relax
  \def\bibfnamefont#1{#1}\fi
\expandafter\ifx\csname citenamefont\endcsname\relax
  \def\citenamefont#1{#1}\fi
\expandafter\ifx\csname url\endcsname\relax
  \def\url#1{\texttt{#1}}\fi
\expandafter\ifx\csname urlprefix\endcsname\relax\def\urlprefix{URL }\fi
\providecommand{\bibinfo}[2]{#2}
\providecommand{\eprint}[2][]{\url{#2}}

\bibitem[{\citenamefont{Herbst}(1991)}]{Herbst1991}
\bibinfo{author}{\bibfnamefont{J.~F.} \bibnamefont{Herbst}},
  \bibinfo{journal}{Rev. Mod. Phys.} \textbf{\bibinfo{volume}{63}},
  \bibinfo{pages}{819} (\bibinfo{year}{1991}).

\bibitem[{\citenamefont{Sagawa et~al.}(1984)\citenamefont{Sagawa, Fujimura,
  Togawa, Yamamoto, and Matsuura}}]{Sagawa1984}
\bibinfo{author}{\bibfnamefont{M.}~\bibnamefont{Sagawa}},
  \bibinfo{author}{\bibfnamefont{S.}~\bibnamefont{Fujimura}},
  \bibinfo{author}{\bibfnamefont{N.}~\bibnamefont{Togawa}},
  \bibinfo{author}{\bibfnamefont{H.}~\bibnamefont{Yamamoto}}, \bibnamefont{and}
  \bibinfo{author}{\bibfnamefont{Y.}~\bibnamefont{Matsuura}},
  \bibinfo{journal}{J. Appl. Phys.} \textbf{\bibinfo{volume}{55}},
  \bibinfo{pages}{2083} (\bibinfo{year}{1984}).

\bibitem[{\citenamefont{Croat et~al.}(1984)\citenamefont{Croat, Herbst, Lee,
  and Pinkerton}}]{Croat1984}
\bibinfo{author}{\bibfnamefont{J.~J.} \bibnamefont{Croat}},
  \bibinfo{author}{\bibfnamefont{J.~F.} \bibnamefont{Herbst}},
  \bibinfo{author}{\bibfnamefont{R.~W.} \bibnamefont{Lee}}, \bibnamefont{and}
  \bibinfo{author}{\bibfnamefont{F.~E.} \bibnamefont{Pinkerton}},
  \bibinfo{journal}{J. Appl. Phys.} \textbf{\bibinfo{volume}{55}},
  \bibinfo{pages}{2078} (\bibinfo{year}{1984}).

\bibitem[{\citenamefont{Vial et~al.}(2002)\citenamefont{Vial, Joly, Nevalainen,
  Sagawa, Hiraga, and Park}}]{Vial2002}
\bibinfo{author}{\bibfnamefont{F.}~\bibnamefont{Vial}},
  \bibinfo{author}{\bibfnamefont{F.}~\bibnamefont{Joly}},
  \bibinfo{author}{\bibfnamefont{E.}~\bibnamefont{Nevalainen}},
  \bibinfo{author}{\bibfnamefont{M.}~\bibnamefont{Sagawa}},
  \bibinfo{author}{\bibfnamefont{K.}~\bibnamefont{Hiraga}}, \bibnamefont{and}
  \bibinfo{author}{\bibfnamefont{K.~T.} \bibnamefont{Park}},
  \bibinfo{journal}{J. Magn. Magn. Mater.} \textbf{\bibinfo{volume}{242-245}},
  \bibinfo{pages}{1329} (\bibinfo{year}{2002}).

\bibitem[{\citenamefont{Li et~al.}(2009)\citenamefont{Li, Ohkubo, and
  Hono}}]{Li2009}
\bibinfo{author}{\bibfnamefont{W.~F.} \bibnamefont{Li}},
  \bibinfo{author}{\bibfnamefont{T.}~\bibnamefont{Ohkubo}}, \bibnamefont{and}
  \bibinfo{author}{\bibfnamefont{K.}~\bibnamefont{Hono}},
  \bibinfo{journal}{Acta Mater.} \textbf{\bibinfo{volume}{57}},
  \bibinfo{pages}{1337} (\bibinfo{year}{2009}).

\bibitem[{\citenamefont{Toga et~al.}(2016)\citenamefont{Toga, Matsumoto,
  Miyashita, Akai, Doi, Miyake, and Sakuma}}]{Toga2016}
\bibinfo{author}{\bibfnamefont{Y.}~\bibnamefont{Toga}},
  \bibinfo{author}{\bibfnamefont{M.}~\bibnamefont{Matsumoto}},
  \bibinfo{author}{\bibfnamefont{S.}~\bibnamefont{Miyashita}},
  \bibinfo{author}{\bibfnamefont{H.}~\bibnamefont{Akai}},
  \bibinfo{author}{\bibfnamefont{S.}~\bibnamefont{Doi}},
  \bibinfo{author}{\bibfnamefont{T.}~\bibnamefont{Miyake}}, \bibnamefont{and}
  \bibinfo{author}{\bibfnamefont{A.}~\bibnamefont{Sakuma}},
  \bibinfo{journal}{Phys. Rev. B} \textbf{\bibinfo{volume}{94}},
  \bibinfo{pages}{174433} (\bibinfo{year}{2016}).

\bibitem[{\citenamefont{Nishino et~al.}(2017)\citenamefont{Nishino, Toga,
  Miyashita, Akai, Sakuma, and Hirosawa}}]{Nishino2017}
\bibinfo{author}{\bibfnamefont{M.}~\bibnamefont{Nishino}},
  \bibinfo{author}{\bibfnamefont{Y.}~\bibnamefont{Toga}},
  \bibinfo{author}{\bibfnamefont{S.}~\bibnamefont{Miyashita}},
  \bibinfo{author}{\bibfnamefont{H.}~\bibnamefont{Akai}},
  \bibinfo{author}{\bibfnamefont{A.}~\bibnamefont{Sakuma}}, \bibnamefont{and}
  \bibinfo{author}{\bibfnamefont{S.}~\bibnamefont{Hirosawa}},
  \bibinfo{journal}{Phys. Rev. B} \textbf{\bibinfo{volume}{95}},
  \bibinfo{pages}{094429} (\bibinfo{year}{2017}).

\bibitem[{\citenamefont{Miura et~al.}(2014)\citenamefont{Miura, Tsuchiura, and
  Yoshioka}}]{Miura2014}
\bibinfo{author}{\bibfnamefont{Y.}~\bibnamefont{Miura}},
  \bibinfo{author}{\bibfnamefont{H.}~\bibnamefont{Tsuchiura}},
  \bibnamefont{and} \bibinfo{author}{\bibfnamefont{T.}~\bibnamefont{Yoshioka}},
  \bibinfo{journal}{J. Appl. Phys.} \textbf{\bibinfo{volume}{115}},
  \bibinfo{pages}{17A765} (\bibinfo{year}{2014}).

\bibitem[{\citenamefont{Yamada et~al.}(1988)\citenamefont{Yamada, Kato,
  Yamamoto, and Nakagawa}}]{Yamada1988}
\bibinfo{author}{\bibfnamefont{M.}~\bibnamefont{Yamada}},
  \bibinfo{author}{\bibfnamefont{H.}~\bibnamefont{Kato}},
  \bibinfo{author}{\bibfnamefont{H.}~\bibnamefont{Yamamoto}}, \bibnamefont{and}
  \bibinfo{author}{\bibfnamefont{Y.}~\bibnamefont{Nakagawa}},
  \bibinfo{journal}{Phys. Rev. B} \textbf{\bibinfo{volume}{38}},
  \bibinfo{pages}{620} (\bibinfo{year}{1988}).

\bibitem[{\citenamefont{Hirosawa et~al.}(2017)\citenamefont{Hirosawa, Nishino,
  and Miyashita}}]{Hirosawa2017}
\bibinfo{author}{\bibfnamefont{S.}~\bibnamefont{Hirosawa}},
  \bibinfo{author}{\bibfnamefont{M.}~\bibnamefont{Nishino}}, \bibnamefont{and}
  \bibinfo{author}{\bibfnamefont{S.}~\bibnamefont{Miyashita}},
  \bibinfo{journal}{Adv. Nat. Sci. Nanosci. Nanotechnol.}
  \textbf{\bibinfo{volume}{8}}, \bibinfo{pages}{013002} (\bibinfo{year}{2017}).

\bibitem[{\citenamefont{Skomski}(1998)}]{Skomski1998a}
\bibinfo{author}{\bibfnamefont{R.}~\bibnamefont{Skomski}}, \bibinfo{journal}{J.
  Appl. Phys.} \textbf{\bibinfo{volume}{83}}, \bibinfo{pages}{6724}
  (\bibinfo{year}{1998}).

\bibitem[{\citenamefont{Sasaki et~al.}(2015)\citenamefont{Sasaki, Miura, and
  Sakuma}}]{Sasaki2015}
\bibinfo{author}{\bibfnamefont{R.}~\bibnamefont{Sasaki}},
  \bibinfo{author}{\bibfnamefont{D.}~\bibnamefont{Miura}}, \bibnamefont{and}
  \bibinfo{author}{\bibfnamefont{A.}~\bibnamefont{Sakuma}},
  \bibinfo{journal}{Appl. Phys. Express} \textbf{\bibinfo{volume}{8}},
  \bibinfo{pages}{043004} (\bibinfo{year}{2015}).

\bibitem[{\citenamefont{Frei et~al.}(1957)\citenamefont{Frei, Shtrikman, and
  Treves}}]{Frei1957}
\bibinfo{author}{\bibfnamefont{E.~H.} \bibnamefont{Frei}},
  \bibinfo{author}{\bibfnamefont{S.}~\bibnamefont{Shtrikman}},
  \bibnamefont{and} \bibinfo{author}{\bibfnamefont{D.}~\bibnamefont{Treves}},
  \bibinfo{journal}{Phys. Rev.} \textbf{\bibinfo{volume}{106}},
  \bibinfo{pages}{446} (\bibinfo{year}{1957}).

\bibitem[{\citenamefont{Booth et~al.}(1995)\citenamefont{Booth, MacIsaac,
  Whitehead, and De'Bell}}]{Booth1995}
\bibinfo{author}{\bibfnamefont{I.}~\bibnamefont{Booth}},
  \bibinfo{author}{\bibfnamefont{A.~B.} \bibnamefont{MacIsaac}},
  \bibinfo{author}{\bibfnamefont{J.~P.} \bibnamefont{Whitehead}},
  \bibnamefont{and} \bibinfo{author}{\bibfnamefont{K.}~\bibnamefont{De'Bell}},
  \bibinfo{journal}{Phys. Rev. Lett.} \textbf{\bibinfo{volume}{75}},
  \bibinfo{pages}{950} (\bibinfo{year}{1995}).

\bibitem[{\citenamefont{Iwano et~al.}(2015)\citenamefont{Iwano, Mitsumata, and
  Ono}}]{Iwano2015}
\bibinfo{author}{\bibfnamefont{K.}~\bibnamefont{Iwano}},
  \bibinfo{author}{\bibfnamefont{C.}~\bibnamefont{Mitsumata}},
  \bibnamefont{and} \bibinfo{author}{\bibfnamefont{K.}~\bibnamefont{Ono}},
  \bibinfo{journal}{J. Appl. Phys.} \textbf{\bibinfo{volume}{117}},
  \bibinfo{pages}{17A704} (\bibinfo{year}{2015}).

\bibitem[{\citenamefont{Appel}(1985)}]{Appel1985}
\bibinfo{author}{\bibfnamefont{A.~W.} \bibnamefont{Appel}},
  \bibinfo{journal}{SIAM J. ScI. STAT. Comput} \textbf{\bibinfo{volume}{6}},
  \bibinfo{pages}{85} (\bibinfo{year}{1985}).

\bibitem[{\citenamefont{Barnes and Hut}(1986)}]{Barnes1986}
\bibinfo{author}{\bibfnamefont{J.}~\bibnamefont{Barnes}} \bibnamefont{and}
  \bibinfo{author}{\bibfnamefont{P.}~\bibnamefont{Hut}},
  \bibinfo{journal}{Nature} \textbf{\bibinfo{volume}{324}},
  \bibinfo{pages}{446} (\bibinfo{year}{1986}).

\bibitem[{\citenamefont{Carrier et~al.}(1988)\citenamefont{Carrier, Greengard,
  and Rokhlin}}]{Carrier1988}
\bibinfo{author}{\bibfnamefont{J.}~\bibnamefont{Carrier}},
  \bibinfo{author}{\bibfnamefont{L.}~\bibnamefont{Greengard}},
  \bibnamefont{and} \bibinfo{author}{\bibfnamefont{V.}~\bibnamefont{Rokhlin}},
  \bibinfo{journal}{SIAM J. Sci. Stat. Comput.} \textbf{\bibinfo{volume}{9}},
  \bibinfo{pages}{669} (\bibinfo{year}{1988}).

\bibitem[{\citenamefont{Makino}(1990)}]{Makino1990}
\bibinfo{author}{\bibfnamefont{J.}~\bibnamefont{Makino}}, \bibinfo{journal}{J.
  Comput. Phys.} \textbf{\bibinfo{volume}{87}}, \bibinfo{pages}{148}
  (\bibinfo{year}{1990}).

\bibitem[{\citenamefont{Darden et~al.}(1993)\citenamefont{Darden, York, and
  Pedersen}}]{Darden1993}
\bibinfo{author}{\bibfnamefont{T.}~\bibnamefont{Darden}},
  \bibinfo{author}{\bibfnamefont{D.}~\bibnamefont{York}}, \bibnamefont{and}
  \bibinfo{author}{\bibfnamefont{L.}~\bibnamefont{Pedersen}},
  \bibinfo{journal}{J. Chem. Phys.} \textbf{\bibinfo{volume}{98}},
  \bibinfo{pages}{10089} (\bibinfo{year}{1993}).

\bibitem[{\citenamefont{Essmann et~al.}(1995)\citenamefont{Essmann, Perera,
  Berkowitz, Darden, Lee, and Pedersen}}]{Essmann1995}
\bibinfo{author}{\bibfnamefont{U.}~\bibnamefont{Essmann}},
  \bibinfo{author}{\bibfnamefont{L.}~\bibnamefont{Perera}},
  \bibinfo{author}{\bibfnamefont{M.~L.} \bibnamefont{Berkowitz}},
  \bibinfo{author}{\bibfnamefont{T.}~\bibnamefont{Darden}},
  \bibinfo{author}{\bibfnamefont{H.}~\bibnamefont{Lee}}, \bibnamefont{and}
  \bibinfo{author}{\bibfnamefont{L.~G.} \bibnamefont{Pedersen}},
  \bibinfo{journal}{J. Chem. Phys.} \textbf{\bibinfo{volume}{103}},
  \bibinfo{pages}{8577} (\bibinfo{year}{1995}).

\bibitem[{\citenamefont{Luijten and BL{\"{O}}TE}(1995)}]{LUIJTEN1995}
\bibinfo{author}{\bibfnamefont{E.}~\bibnamefont{Luijten}} \bibnamefont{and}
  \bibinfo{author}{\bibfnamefont{H.~W.} \bibnamefont{Bl{\"{o}}te}},
  \bibinfo{journal}{Int. J. Mod. Phys. C} \textbf{\bibinfo{volume}{06}},
  \bibinfo{pages}{359} (\bibinfo{year}{1995}).

\bibitem[{\citenamefont{Hinzke and Nowak}(2000)}]{Hinzke2000}
\bibinfo{author}{\bibfnamefont{D.}~\bibnamefont{Hinzke}} \bibnamefont{and}
  \bibinfo{author}{\bibfnamefont{U.}~\bibnamefont{Nowak}}, \bibinfo{journal}{J.
  Magn. Magn. Mater.} \textbf{\bibinfo{volume}{221}}, \bibinfo{pages}{365}
  (\bibinfo{year}{2000}).

\bibitem[{\citenamefont{Sasaki and Matsubara}(2008)}]{Sasaki2008}
\bibinfo{author}{\bibfnamefont{M.}~\bibnamefont{Sasaki}} \bibnamefont{and}
  \bibinfo{author}{\bibfnamefont{F.}~\bibnamefont{Matsubara}},
  \bibinfo{journal}{J. Phys. Soc. Jpn.} \textbf{\bibinfo{volume}{77}},
  \bibinfo{pages}{024004} (\bibinfo{year}{2008}).

\bibitem[{\citenamefont{Sasaki}(2010)}]{Sasaki2010}
\bibinfo{author}{\bibfnamefont{M.}~\bibnamefont{Sasaki}},
  \bibinfo{journal}{Phys. Rev. E} \textbf{\bibinfo{volume}{82}},
  \bibinfo{pages}{031118} (\bibinfo{year}{2010}).

\bibitem[{\citenamefont{Fukui and Todo}(2009)}]{Fukui2009}
\bibinfo{author}{\bibfnamefont{K.}~\bibnamefont{Fukui}} \bibnamefont{and}
  \bibinfo{author}{\bibfnamefont{S.}~\bibnamefont{Todo}}, \bibinfo{journal}{J.
  Comput. Phys.} \textbf{\bibinfo{volume}{228}}, \bibinfo{pages}{2629}
  (\bibinfo{year}{2009}).

\bibitem[{\citenamefont{Endo et~al.}(2015)\citenamefont{Endo, Toga, and
  Sasaki}}]{Endo2015a}
\bibinfo{author}{\bibfnamefont{E.}~\bibnamefont{Endo}},
  \bibinfo{author}{\bibfnamefont{Y.}~\bibnamefont{Toga}}, \bibnamefont{and}
  \bibinfo{author}{\bibfnamefont{M.}~\bibnamefont{Sasaki}},
  \bibinfo{journal}{J. Phys. Soc. Jpn.} \textbf{\bibinfo{volume}{84}},
  \bibinfo{pages}{074002} (\bibinfo{year}{2015}).

\bibitem[{\citenamefont{Mak}(2005)}]{Mak2005a}
\bibinfo{author}{\bibfnamefont{C.~H.} \bibnamefont{Mak}}, \bibinfo{journal}{J.
  Chem. Phys.} \textbf{\bibinfo{volume}{122}} (\bibinfo{year}{2005}).

\bibitem[{\citenamefont{Momma and Izumi}(2011)}]{Momma2011}
\bibinfo{author}{\bibfnamefont{K.}~\bibnamefont{Momma}} \bibnamefont{and}
  \bibinfo{author}{\bibfnamefont{F.}~\bibnamefont{Izumi}}, \bibinfo{journal}{J.
  Appl. Crystallogr.} \textbf{\bibinfo{volume}{44}}, \bibinfo{pages}{1272}
  (\bibinfo{year}{2011}).

\bibitem[{\citenamefont{Liechtenstein et~al.}(1987)\citenamefont{Liechtenstein,
  Katsnelson, Antropov, and Gubanov}}]{Liechtenstein1987}
\bibinfo{author}{\bibfnamefont{A.~I.} \bibnamefont{Liechtenstein}},
  \bibinfo{author}{\bibfnamefont{M.~I.} \bibnamefont{Katsnelson}},
  \bibinfo{author}{\bibfnamefont{V.~P.} \bibnamefont{Antropov}},
  \bibnamefont{and} \bibinfo{author}{\bibfnamefont{V.~A.}
  \bibnamefont{Gubanov}}, \bibinfo{journal}{J. Magn. Magn. Mater.}
  \textbf{\bibinfo{volume}{67}}, \bibinfo{pages}{65} (\bibinfo{year}{1987}).

\end{thebibliography}
\end{document}